# Career-Aware Resume Tailoring via Multi-Source Retrieval-Augmented Generation with Provenance Tracking: A Case Study

**Kumar Abhinav**

rushtoabhinavin@gmail.com

***Abstract—***

*AI-assisted resume tailoring systems commonly operate on a single uploaded resume, which limits their ability to recover relevant experience omitted from the current draft and makes it difficult for users to distinguish grounded edits from model-generated suggestions. This paper presents Resume Tailor, an agentic resume-tailoring system that maintains a longitudinal career vault in a vector database and uses multi-source retrieval-augmented generation (RAG) to assemble job-specific resume content from historical resumes and structured career records. The system is implemented as a 12-node LangGraph pipeline with typed state management, hybrid semantic-lexical confidence scoring, provenance-aware fallback generation, anti-hallucination guardrails, and a conditional review loop. We report a pilot evaluation on nine job descriptions (JDs) across software engineering, data analytics, and business analysis roles using a single candidate's career history. For six JDs where the candidate held at least one prior role in the same occupational category, enabling the career vault improved Applicant Tracking System (ATS)-style fit scores by an average of 7.8 points. For two JDs requiring domain-specific expertise absent from the vault, scores decreased by an average of 8.0 points. One partially overlapping role showed a modest gain of 2 points. These results suggest that longitudinal retrieval can improve resume tailoring when relevant prior experience exists, while also highlighting the need for confidence-gated retrieval when domain overlap is weak.*

## I. INTRODUCTION

Applicant Tracking Systems (ATS) shape how resumes are screened before human review. In practice, candidates often adapt the same base resume across multiple applications, emphasizing different experiences depending on role, industry, and required skills. This process is time-consuming and error-prone, and it is increasingly supported by AI-assisted writing tools.

Most current AI resume-tailoring workflows treat each request as a single-document task: the system rewrites only the uploaded resume for the current application. That assumption is restrictive because a candidate's relevant evidence is often distributed across older resume versions, archived application materials, and structured career records. As a result, useful experience can remain inaccessible even when it is genuinely part of the candidate's history.

This limitation motivates three design requirements. First, a tailoring system should recover relevant experience across time rather than only from the active resume draft. Second, it should separate retrieved, user-grounded content from newly generated suggestions so that candidates can review outputs responsibly. Third, it should retain useful refinements across sessions rather than restarting from scratch each time.

To address these requirements, we present Resume Tailor, a career-aware resume-tailoring system built around three ideas: (a) a longitudinal career vault implemented with persistent vector collections over historical resumes and structured career records; (b) provenance-aware generation in which retrieved content and fallback-generated content are tracked separately within the pipeline; and (c) a feedback loop that persists vetted generated content for reuse in later tailoring sessions.

The system is implemented as a 12-node LangGraph pipeline with typed state, hybrid retrieval scoring, anti-hallucination guardrails, and a conditional holistic review step. We evaluate the approach in a pilot case study using nine real JDs and show where longitudinal retrieval helps, where it hurts, and why those boundary conditions matter.

## II. RELATED WORK

### *A. Resume-Job Matching*

Automated matching between resumes and job descriptions has evolved from lexical similarity methods to representation learning and transformer-based architectures. Earlier work studied recommendation-based job matching and bilateral matching behavior [1]. Representation-learning methods based on distributed embeddings [8] and sentence-level transformer embeddings [7] helped motivate later matching systems. More recent work applies contextual transformer models to resume-job matching and competence prediction [2], while recent research has also examined the extent to which LLM-based resume assessments align with human evaluation [3].

### B. Explainable Hiring Pipelines

Recent hiring-oriented NLP systems emphasize structured extraction, semantic matching, and explainability. The Smart-Hiring pipeline [4] combines resume parsing, entity extraction, and semantic matching in an end-to-end system. These systems focus primarily on ranking or matching candidates rather than rewriting resume content for a specific application.

### C. Retrieval-Augmented Generation

Retrieval-augmented generation (RAG) combines external retrieval with language-model generation [5]. In contrast to open-domain RAG, our setting uses a private corpus consisting of a candidate's own professional history. This design shifts the primary objective from knowledge access to grounded evidence recovery across time.

### D. Agentic and Iterative AI Workflows

LangGraph [6] provides a graph-oriented framework for building stateful, multi-step agent workflows with conditional routing and persistence. Our system uses this orchestration style to coordinate retrieval, fallback generation, review, and persistence in a single pipeline.

## III. SYSTEM ARCHITECTURE

### A. Overview

Resume Tailor implements a 12-node pipeline with one conditional feedback edge. Fig. 1 illustrates the complete architecture. The nodes are organized into five functional stages: (1) ingestion and parsing, (2) retrieval and fallback, (3) content generation, (4) assembly and review, and (5) scoring and output. If the final review stage identifies major quality issues, the graph routes back to the rewrite stage for one additional refinement pass.

### B. Career Vault

The career vault is implemented as persistent ChromaDB collections using text-embedding-3-small embeddings (1536 dimensions). The vault stores three categories of information: historical resume chunks parsed from uploaded PDF and DOCX files, structured career-history chunks parsed from CSV or XML records, and previously generated resume chunks that have been retained for reuse. Each stored chunk includes metadata such as section, chunk level, parent chunk identifier, and source-document reference. The chunk schema is source-agnostic, so the same vault design generalizes to other longitudinal career artifacts such as midyear or annual performance reviews, self-appraisals, and, for early-career candidates without formal work history, coursework, capstone projects, and graduate research summaries. Evaluating these additional sources is left to future work, but the architecture does not require modification to accommodate them.

### C. Hybrid Confidence Scoring

Retrieved snippets receive a hybrid confidence score:

$$confidence = \alpha \times semantic + (1 - \alpha) \times lexical$$

where semantic is a normalized cosine similarity score, lexical is the mean fuzzy overlap (via rapidfuzz partial_ratio) with extracted JD elements, and $\alpha = 0.6$ by default. The default $\alpha = 0.6$ slightly favors semantic similarity over lexical overlap, reflecting the observation that resume–JD matching benefits from semantic generalization (e.g., recognizing "led a cross-functional initiative" as evidence for a "stakeholder management" requirement) while still rewarding exact matches on high-signal tokens such as specific technologies, certifications, and framework names via the lexical component. This value was selected empirically during development rather than through a formal hyperparameter search, and $\alpha$ is exposed as a configuration parameter; a systematic ablation over $\alpha \in [0, 1]$ is left to future work. Snippets below a configurable threshold (default 0.75) are filtered before rewriting. We note that this is a hybrid scoring function applied to candidates returned by dense retrieval, not a true hybrid retriever in which a sparse method such as BM25 and a dense method independently return top-k candidates that are then fused (for example, via reciprocal rank fusion). A sparse-plus-dense hybrid retriever may surface candidates missed by embeddings alone—particularly for exact-match tokens such as tool names, framework versions, and proprietary system acronyms that are common in JDs—and comparing the two configurations is left to future work.

### D. Provenance-Aware Fallback

When retrieval yields insufficient evidence, the system activates a three-tier fallback strategy. Tier 1 uses high-confidence vault snippets labeled with their source collection and document reference. Tier 2 uses language-model generation with prompts that explicitly prohibit fabricated employers, roles, or performance metrics. Tier 3 uses deterministic templates derived from JD elements.

The pipeline tracks retrieval mode and keeps fallback outputs separate from vault-grounded experience content. Critically, fallback-generated content (Tiers 2 and 3) is never merged into employer-specific experience entries. Only vault-retrieved content is eligible for injection into company sections, preventing attribution of fabricated accomplishments to real employers. Fallback content that survives the review stages is available as internal tailored_highlights metadata for the candidate's inspection, but it is excluded from the rendered PDF resume to avoid presenting unverifiable claims as part of the candidate's work history.

### E. Quality Assurance

Three stages contribute to quality control. A guardrails stage screens for unsupported claims, fabricated company names, and formatting problems. A polish stage reorganizes and normalizes content across entries, including ATS Unicode normalization that converts smart quotes, em-dashes, and ligatures to their ASCII equivalents. A holistic review stage evaluates coherence, duplication, JD alignment, and unnatural AI-written phrasing, and can trigger one additional rewrite pass through the conditional feedback edge.

### F. ATS-Oriented Scoring

To support comparative analysis, the system computes ATS-style fit scores under five weighting profiles. These profiles do not claim to reproduce proprietary ATS implementations; instead, they provide a controlled scoring framework for relative comparison across experimental conditions. Table I lists the exact weight configurations.

TABLE I. ATS SCORING PROFILE WEIGHTS

| Profile | Skill | Resp. | Qual. | Best |
|---|---|---|---|---|
| Skills-Heavy | 0.50 | 0.30 | 0.20 | max |
| Role-Aligned | 0.40 | 0.40 | 0.20 | max |
| Resp.-First | 0.30 | 0.50 | 0.20 | max |
| Qual.-Heavy | 0.25 | 0.25 | 0.50 | max |
| Balanced | 0.33 | 0.34 | 0.33 | max |

The reported Overall Fit score is the mean across all five profiles for each run. The system also computes a Best-Profile score (the maximum), but Table IV reports the mean for consistency with the dashboard display.

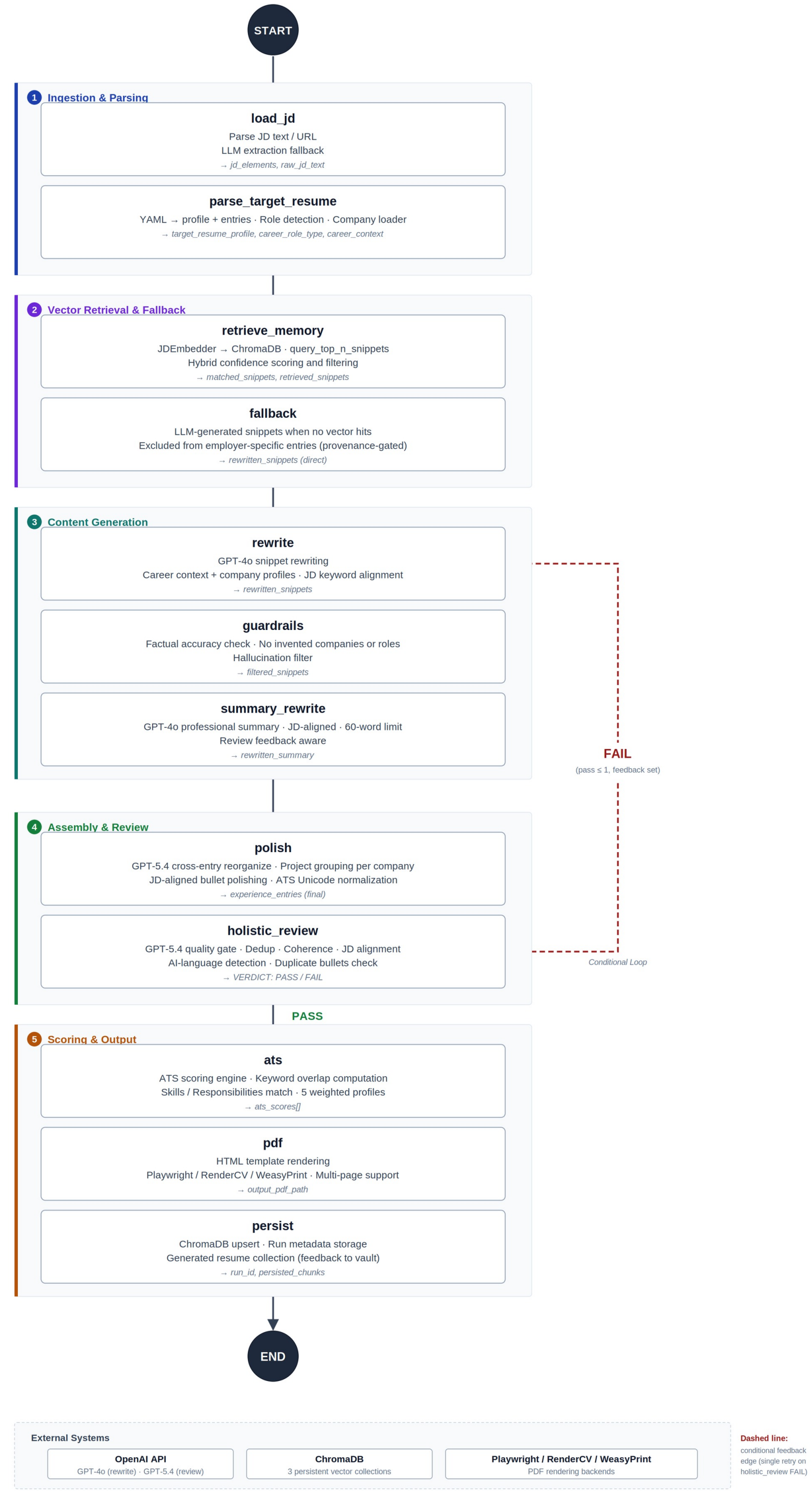

*Fig. 1. Resume Tailor system architecture: 12-node LangGraph pipeline with conditional feedback loop. Nodes are grouped into five stages with external dependencies (OpenAI API, Playwright, ChromaDB) shown at bottom.*

## IV. IMPLEMENTATION

The system is implemented in Python 3.11 as a FastAPI service totaling approximately 11,200 lines across 75 modules. The API exposes endpoints for career-vault indexing, resume tailoring, PDF generation, and run-history inspection. Generated resume content can be persisted back to the vault, enabling reuse in later sessions. The system is containerized via Docker with all dependencies pre-installed.

TABLE II. TECHNOLOGY STACK

| Component | Technology |
|---|---|
| Orchestration | LangGraph, optional Temporal |
| Vector Store | ChromaDB (3 collections, 1536-dim) |
| Embeddings | OpenAI text-embedding-3-small |
| LLMs | GPT-4o (rewrite/summary), GPT-5.4 (review) |
| NLP | spaCy en_core_web_sm, rapidfuzz |
| PDF | Playwright ATS Pro, WeasyPrint, RenderCV |
| API | FastAPI, Pydantic v2, SQLite |
| Observability | LangSmith tracing, structured logging |

## V. EVALUATION

### *A. Experimental Design*

We conducted a pilot evaluation using nine JDs collected from active postings between February and April 2026 across three role families. Table III identifies each JD by employer, role title, and posting date. The career vault was built from five historical resume versions spanning approximately fifteen years of experience for a single candidate, covering software engineering, data analytics, supply chain optimization, and AI product development.

### *B. Conditions*

Each JD was evaluated under two conditions:

Baseline: vector retrieval disabled. The system operated on a single target resume (a current software-engineering-focused resume dated March 2026) while keeping all downstream polish, review, scoring, and Playwright ATS Pro rendering stages unchanged. Although the fallback node generates LLM-written snippets when retrieval is absent, those snippets are not consumed by the ATS scoring or PDF rendering stages; the scoring and rendering logic gates on vault-matched content. The baseline therefore evaluates how the pipeline scores and renders the candidate's existing resume content without vault augmentation.

Career Vault: vector retrieval enabled across all five indexed historical resume versions and structured career records.

All reported PDFs in this evaluation were rendered with the Playwright ATS Pro template. All other configuration settings—including LLM model, confidence threshold, ATS profile weights, and PDF template—were held constant across the two conditions.

### *C. Grouping Criteria*

To organize the results, we classify each JD into one of three categories based on the occupational overlap between the JD's primary role family and the candidate's indexed career history. A JD is classified as domain-aligned if the candidate held at least one prior role in the same occupational category (e.g., a Data Analyst JD when the vault contains prior analytics roles). A JD is classified as domain-distant if the JD's primary skill requirements fall in a specialized domain (e.g., bioinformatics, neuroscience) for which the vault contains no relevant experience. A JD is classified as partially aligned if it shares the candidate's broad occupational field but requires specific technologies or frameworks not represented in the vault. These categories were assigned before examining the vault-vs-baseline score deltas to avoid post-hoc grouping.

### *D. Results*

TABLE III. JOB DESCRIPTIONS USED IN EVALUATION

| Employer | Role Title | Cat. | Posted |
|---|---|---|---|
| Stanford Univ. | Full Stack Developer | SWE | Feb 2026 |
| Stanford Univ. | Sr SWE (Genomics) | SWE | Mar 2026 |
| Dartmouth College | Research SWE | SWE | Mar 2026 |
| Deloitte | Lead OpenAI FDE | SWE | Mar 2026 |
| Deloitte | Data Analyst | DA | Apr 2026 |
| TEKsystems | Data Analyst | DA | Apr 2026 |
| Deloitte | Sr Data Analyst | DA | Apr 2026 |
| UnionWare | Business Analyst | BA | Mar 2026 |
| Manitoba Hydro | SAP BSA | BA | Apr 2026 |

Table IV reports ATS-style overall fit scores under the two conditions. Per-JD scores are the primary result; grouped averages appear in Section VI.

TABLE IV. ATS-STYLE OVERALL FIT SCORES: BASELINE VS. CAREER VAULT

| Job Description | Group | Base | Vault | Δ | Verdict |
|---|---|---|---|---|---|
| Stanford Full Stack | Part. | 53 | 55 | +2 | Partial |
| Stanford Genomics | Dist. | 57 | 49 | −8 | Partial |
| Dartmouth RSE | Dist. | 57 | 49 | −8 | Partial |
| Deloitte OpenAI FDE | Align | 72 | 79 | +7 | Strong |
| Deloitte Data Analyst | Align | 55 | 59 | +4 | Comp. |
| TEK Data Analyst | Align | 80 | 82 | +2 | Strong |
| Deloitte Sr DA | Align | 59 | 82 | +23 | Strong |
| UnionWare BA | Align | 79 | 83 | +4 | Strong |
| SAP BSA | Align | 77 | 84 | +7 | Strong |

### *E. Per-JD Observations*

The largest gain occurred for the Deloitte Senior Data Analyst (+23 points, 59 to 82), where the vault retrieved early-career experience in SAP analytics, Tableau dashboarding, and executive reporting that was absent from the current resume. The SAP BSA (+7) and Deloitte OpenAI FDE (+7) showed similar patterns: the vault surfaced relevant but de-emphasized experience. The TEK Data Analyst (+2) and Deloitte Data Analyst (+4) showed smaller gains because the current resume already covered much of the required skill set.

The two domain-distant JDs—Stanford Genomics and Dartmouth RSE—both decreased by 8 points. These roles require bioinformatics tools (SAMtools, BEDtools, genomic data formats) and neuroscience standards (BIDS, NWB) respectively, none of which exist in any form in the vault. The retrieval mechanism surfaced semantically similar but domain-mismatched snippets, reducing specificity compared to the baseline.

The Stanford Full Stack Developer (+2) falls between the two patterns. The candidate has general web development experience, but lacks the specific technologies (Storyblok, NextJS, TailwindCSS, ISR/SSR/SSG) that the JD emphasizes, producing a small but not decisive gain.

### *F. Qualitative Provenance Observations*

Although this evaluation does not directly measure provenance accuracy or longitudinal learning—both of which require human annotation or multi-session studies—we note two qualitative observations from the experimental runs.

TABLE V. QUALITATIVE OBSERVATIONS ON DESIGN FEATURES

| Feature | Observation |
|---|---|
| Anti-hallucination | No company-name fabrication observed in any of 18 runs (9 JDs × 2 conditions). Consistent with prompt constraints and the merge-exclusion policy. |
| Fallback exclusion | Fallback-generated bullets were correctly excluded from employer entries in all vault-enabled runs. Content was available as internal metadata but not rendered in the PDF. |
| Vault reuse | Content persisted from run 1 was retrievable in subsequent runs using the same vault. Longitudinal accumulation effect not measured. |

These observations are not quantitative results and should not be interpreted as validated claims. Formal evaluation of provenance accuracy and longitudinal learning utility remains future work.

## VI. DISCUSSION

### *A. Grouped Averages*

Aggregating the per-JD results by the grouping criteria defined in Section V-C, the six domain-aligned JDs showed a mean improvement of +7.8 points (range: +2 to +23), the two domain-distant JDs showed a mean decrease of −8.0 points, and the single partially aligned JD showed a gain of +2 points. These averages are reported as descriptive summaries; the per-JD scores in Table IV are the primary result.

### *B. What the Career Vault Adds*

The primary contribution of the career vault is improved recall over the candidate's own history. In the strongest cases, the system surfaced older but relevant evidence that would have been unavailable to a single-document workflow. The Deloitte Senior Data Analyst case illustrates this: the current resume emphasized GenAI engineering, but the vault retrieved decade-old Tableau, SAP data mapping, and executive reporting experience that directly addressed the JD's requirements.

### *C. Boundary Conditions*

The same retrieval mechanism can reduce output quality when the candidate lacks domain-specific background for the target role. In those cases, retrieval may introduce superficially similar but strategically unhelpful content. This finding supports confidence-gated retrieval and possibly a retrieval-abstention mode when evidence quality is weak.

### *D. Provenance as a Design Decision*

The provenance-tracking and fallback-exclusion mechanisms are architectural design decisions motivated by the observation that resume editing is a high-stakes application where inaccurate content can misrepresent a candidate's experience. The current implementation tracks provenance internally and enforces merge restrictions, but does not expose a user-facing provenance interface in the generated resume. Evaluating the utility of explicit provenance labeling for end-user trust remains future work.

### *E. Beyond Candidate-Side Audit: The Employer-Signal Asymmetry*

The quality-assurance mechanisms described in Section III-E audit only the candidate side of the application pipeline: they constrain what the system will write about the candidate, based on what the candidate can evidence. They do not audit the job description itself. This is a deliberate scope choice, but it also leaves an important source of matching error unaddressed. Two well-documented problems on the employer side reduce the meaningfulness of any JD-to-resume fit score. First, JDs frequently diverge from the actual requirements of the role, because they are assembled from templates, inherited from previous requisitions, or written by stakeholders who are not the eventual hiring decision-maker. Second, a non-trivial fraction of posted openings are not associated with a real, fundable vacancy—so-called ghost postings—where no candidate would be hired regardless of fit.

Both failure modes are largely invisible from the JD text itself, which is the only signal our pipeline consumes on the employer side. A candidate-optimized resume directed at a misspecified or non-hiring JD receives no value from tailoring, regardless of how grounded the retrieved evidence is. We therefore view employer-side auditing as an orthogonal research problem to the one addressed in this paper, but one that bounds the achievable real-world utility of any resume-tailoring system. Potential signals worth investigating in future work include longitudinal tracking of JD reposting patterns, cross-posting fingerprints that indicate template reuse, divergence between JD text and the skill distribution of the team's recent hires, and time-to-first-interview statistics aggregated across postings from the same employer. None of these provides a definitive label, but in combination they could produce a JD-confidence estimate that complements the candidate-side confidence already exposed by our retrieval layer. We flag this explicitly so that downstream users of fit scores—whether candidates, coaches, or automated ranking systems—understand what our scores do and do not measure.

### *F. Limitations*

This study has several limitations. First, the evaluation is based on one candidate and nine JDs, so the results should be interpreted as descriptive rather than statistically generalizable. Second, the scoring framework uses synthetic weighting profiles rather than proprietary ATS scoring functions. Third, the implementation uses closed-weight models (GPT-4o, GPT-5.4), which can affect reproducibility across time and model versions. Fourth, the pipeline does not preserve the original visual layout of uploaded resumes and depends on document parsing that can be lossy for complex PDFs. Fifth, while the baseline condition approximates a single-document setting, it still uses the same pipeline stages as the vault-enabled condition; a direct comparison with commercial tools was not conducted. Sixth, the JDs used were collected from public postings; future work should include archived copies for full reproducibility.

## VII. CONCLUSION

This paper presented Resume Tailor, a career-aware resume-tailoring system that combines multi-source retrieval, provenance-aware fallback logic, guardrails, and iterative review in a LangGraph-based workflow. In a pilot case study with nine JDs, enabling retrieval across a longitudinal career vault improved ATS-style fit scores for six domain-aligned roles by an average of 7.8 points, while degrading performance by 8.0 points on two domain-distant roles. One partially aligned role showed a modest gain. These findings suggest that career-aware retrieval is useful when grounded prior evidence exists, but should be paired with confidence gating when target roles exceed the candidate's demonstrated history.

Future work includes larger multi-candidate evaluations across diverse career trajectories and seniority levels; human assessment of factuality and usefulness; retrieval-abstention strategies for domain-distant roles; user-facing provenance interfaces; resume-specific embedding or reranking models better suited to career-history data; true sparse-plus-dense hybrid retrieval (for example, BM25 combined with dense embeddings via reciprocal rank fusion) to improve recall on exact-match tokens; extension of the career vault to additional longitudinal sources such as performance reviews, self-appraisals, and—for students and early-career candidates without formal work history—coursework, capstone projects, and research writeups; and, on the employer-signal side discussed in Section VI-E, exploration of JD-confidence estimation using reposting patterns, template-reuse fingerprints, and hiring-outcome statistics as complementary signals to candidate-side retrieval confidence.